\documentclass[twocolumn,prl,aps,showpacs,preprintnumbers,superscriptaddress, amsmath,amssymb,twoside,floatfix]{revtex4-1}

\usepackage{graphicx}		
\usepackage{pstricks}
\usepackage{bm}			
\usepackage{units}		
\usepackage{color}		
\usepackage{ulem}
\usepackage{hyperref}
\usepackage{xspace}

\newcommand{\NdFeBO}{NdFe$_{\mbox{\scriptsize 3}}$(BO$_{\mbox{\scriptsize 3}}$)$_{\mbox{\scriptsize 4}}$\xspace}

\newcommand{\NdLtwo}{Nd\,$L_{\mbox{\scriptsize 2}}$}

\newcommand{\qic}{$\mathbf{q}_{\mbox{\scriptsize ICM}}$}
\newcommand{\qc}{$\mathbf{q}_{\mbox{\scriptsize CM}}$}
\newcommand{\icp}{$(0, 0, 3n+3/2 \pm \delta)$}
\newcommand{\cp}{$(0, 0, 3n+3/2)$}
\newcommand{\Pfe}{$\mathcal{P}$}
\newcommand{\Eval}{$8.8 \cdot \unit[10^4]{V/m}$}
\newcommand{\TN}{$T_\mathrm{N}$}
\newcommand{\Tic}{$T_{\mathrm{IC}}$}
\newcommand{\circr}{$c_\mathrm{r}$}
\newcommand{\cl}{$c_\mathrm{l}$}

\begin{document}

\title{Control of the magnetic phase coexistence in \NdFeBO by electric fields}

\author{S.~Partzsch}

\altaffiliation{Present address: Optotransmitter und Umweltschutz-technologie OUT e.V., K\"openicker Str. 325, Haus 201, D-12555 Berlin, Germany}
\affiliation{Leibniz Institute for Solid State and Materials Research IFW Dresden, Helmholtzstrasse 20, D-01069 Dresden, Germany}

\author{J.-E.~Hamann-Borrero}
\affiliation{Leibniz Institute for Solid State and Materials Research IFW Dresden, Helmholtzstrasse 20, D-01069 Dresden, Germany}

\author{C.~Mazzoli}
\affiliation{European Synchrotron Radiation Facility (ESRF), BP 220, 38043 Grenoble, France}
\affiliation{Dipartimento di Fisica e Unit\`a CNISM, Politecnico di Milano, Piazza Leonardo Da Vinci 32,I-20133 Milano, Italy}

\author{J.~Herrero-Martin}
\affiliation{European Synchrotron Radiation Facility (ESRF), BP 220, 38043 Grenoble, France}
\affiliation{ALBA Synchrotron Light Source, E-08290 Cerdanyola del Vall\`es, Barcelona, Spain}

\author{A.~Vasiliev}
\affiliation{Low Temperature Physics Department, Faculty of Physics, Moscow State University, Moscow, 119992 Russia}

\author{L.~Bezmaternykh}
\affiliation{Low Temperature Physics Department, Faculty of Physics, Moscow State University, Moscow, 119992 Russia}

\author{B.~B\"uchner} 
\affiliation{Leibniz Institute for Solid State and Materials Research IFW Dresden, Helmholtzstrasse 20, D-01069 Dresden, Germany}
\affiliation{Institute for Solid State Physics, Dresden Technical University, TU-Dresden, D-01062 Dresden, Germany}

\author{J.~Geck}
\email{j.geck@ifw-dresden.de}
\affiliation{Leibniz Institute for Solid State and Materials Research IFW Dresden, Helmholtzstrasse 20, D-01069 Dresden, Germany}

\begin{abstract}
We present a resonant x-ray diffraction study of the magnetic order in \NdFeBO\/ and its coupling to applied electric fields. Our high-resolution measurements
reveal a coexistence of two different magnetic phases, which can be triggered effectively by external electric fields. More in detail, the volume fraction of the collinear magnetic phase is found to strongly increase at the expense of helically ordered regions when an electric field is applied.
These results confirm that the collinear magnetic phase is responsible for the ferroelectric polarization of \NdFeBO\ and, more importantly, demonstrate that magnetic phase coexistence provides an alternative route towards materials with a strong magnetoelectric response.
\end{abstract}

\date{Received: \today}
%
\pacs{75.85.+t, 77.80.-e, 64.70.Rh, 61.05.}


\maketitle

Frustrated magnetic materials exhibit a large variety of intriguing physical phenomena, including the concomitant appearance of coupled ferroelectric and magnetic orders\,\cite{Cheong2007,Khomskii2009}. In fact, interlinked magnetic and ferroelectric orders are very rare in nature. Their discovery in variety of frustrated magnets was therefore a surprise and generated a lot of excitement\,\cite{Kimura2003Nat,Hur2004NatureTb}, not only because these phenomena are very interesting from the viewpoint of basic research. 
There is also a significant technological potential\,\cite{Ramesh2007,Ma11}. Especially since a strong magnetoelectric coupling enables to store information in an extremely energy efficient way\,\cite{VazJOP12,Heron14}.  Indeed, earlier experiments could demonstrate modifications of a magnetic order related to the reversal of the electric polarization\,\cite{Radaelli2008Y,Fabrizi2009,Tokunaga2012}. These studies, however, were concerned with electric field induced changes within a single magnetic phase or the manipulation of phase stability close to a magnetic transition\,\cite{BodenthinPRL2008}.

In the present study we consider a different situation, where frustrated magnetic interactions cause two distinct ordered states to be metastable well below the observed critical temperatures, even in the absence of any first order transition. As a specific example, we investigated the electric field dependence of the magnetic order in  \NdFeBO, which is a frustrated magnetic material with a large magnetoelectric response\,\cite{Zvezdin2006}. 
We show that a small perturbation created by an applied {\it electric} field allows controlling the stability of the coexisting collinear and helical {\it magnetic} orders. This  implies that phase coexistence driven by magnetic frustration underlies the magnetoelectric response of \NdFeBO, providing an alternative route towards large magnetoelectric effects.

 \begin{figure}[b!]
\center{
 \includegraphics[width=0.9\columnwidth]{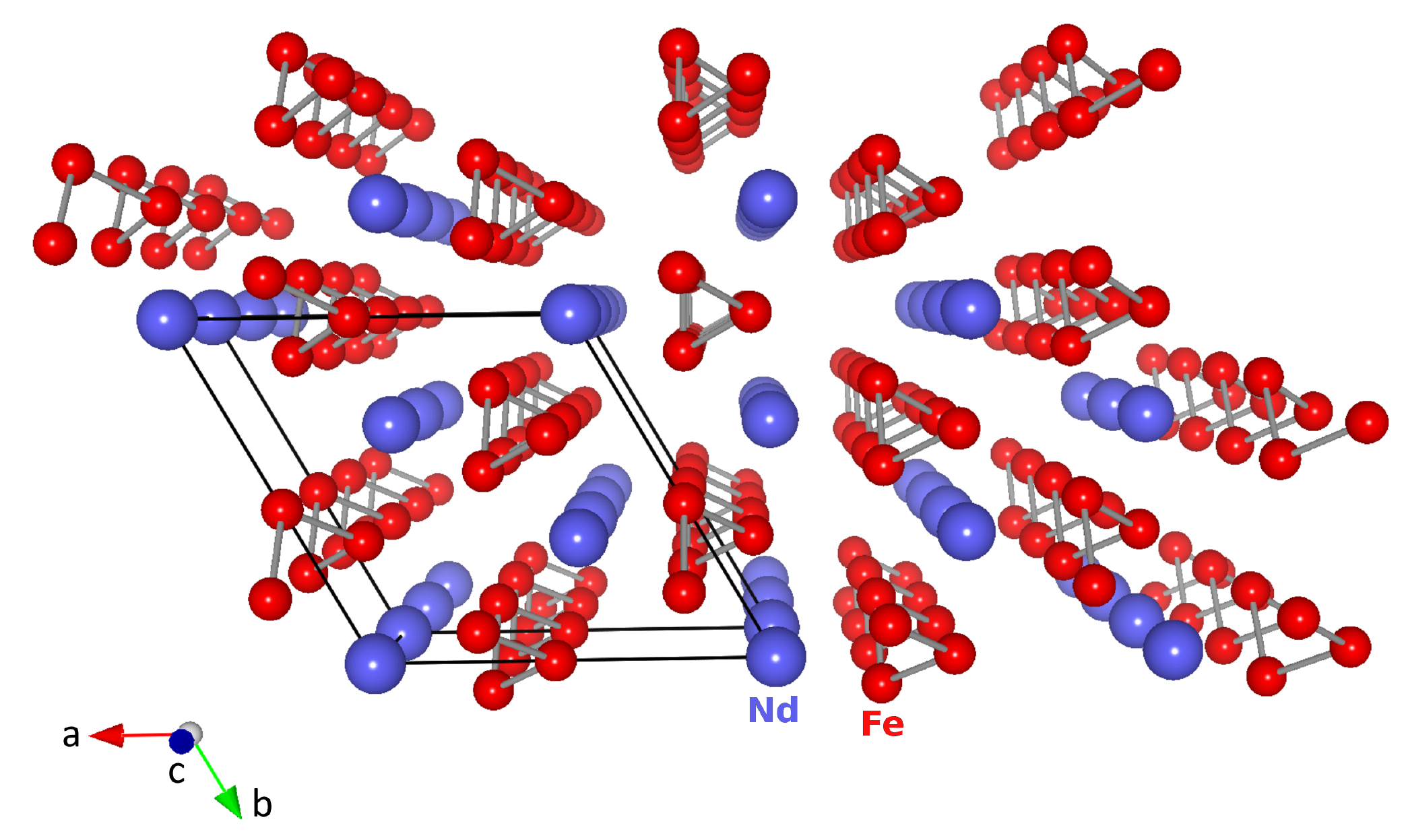}
}
 \caption{(Color online) Crystal structure of \NdFeBO. The material can be described using the rhombohedral space group $R32$ with hexagonal unit cell parameters $a \approx 9.5$\,\AA\ and $c \approx 7.5$\,\AA\/ at room temperature\,\cite{Campa1997}. Only the magnetic sublattices of Fe (red) and Nd (purple) are shown. Fe is coordinated by oxygen octahedra, which form edge sharing helical chains along the $c$-direction. These chains and NdO$_6$ triangular prisms are connected in the $ab$-plane by BO$_3$ triangles (not shown). 
 }
 \label{fig:struc}
 \end{figure}

\begin{figure*}[t!]
\center{
   \includegraphics[width=0.85\textwidth]{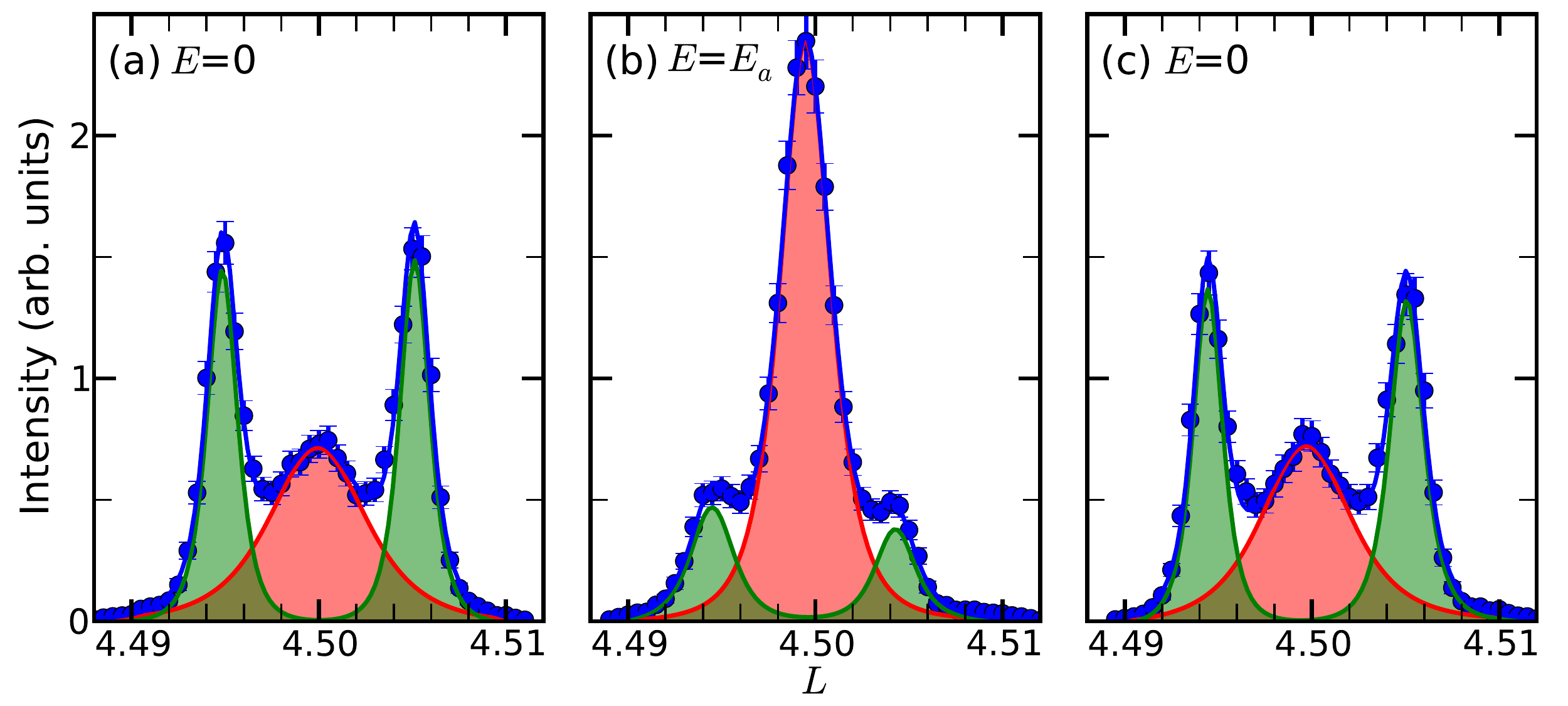}
 }
\caption{(Color online) $L$-scans around the (0 0 4.5) position at $\unit[5.8]{K}$, fitted with three Lorentzian squared peaks and a linear background (continuous lines).
The the sum of the fitted curves agrees well with the experimental data (dots).
The process induced by the $E$-field of $E_a=$ \Eval\ is reversible.
}
\label{fig:E-Leff}
\end{figure*}

The rhombohedral lattice structure of this material is illustrated in Fig.\,\ref{fig:struc}, where, for the sake of simplicity, only the two magnetic sublattices of Fe and Nd are shown. These two sublattices are magnetically coupled and undergo two magnetic phase transitions as a function of temperature\,\cite{Fischer2006,Janoschek2010,HamannBorrero2012HT}: upon cooling, commensurate magnetic (CM) order sets in first at \TN\ $\approx \unit[30]{K}$. 
In this phase the spins are ordered in a collinear fashion, forming ferromagnetic (FM) $ab$-planes, which are coupled antiferromagnetically (AFM) along the $c$-direction. The magnetic modulation vector is commensurate, causing magnetic superlattice peaks at \qc = \cp \/ with integer $n$.
Below \Tic\ $\approx \unit[15]{K}$, a continuous transition into an incommensurate magnetic (ICM) phase occurs, where the FM  moments of the $ab$-planes form helices propagating along the $c$-direction. The helical order is signaled by magnetic superlattice peaks at \qic = \icp, where 
$\delta$ increases continuously with further cooling due to a dramatically decreasing period of the spin helix ranging from $523\times c$ at  \unit[14]{K} down to $146\times c$ at \unit[1.6]{K}. 

Interestingly, 
applying an external magnetic field $B_a \approx \unit[1.3]{T}$ at $\unit[4.8]{K}$ along the $a$-direction induces an electric polarization \Pfe$_a$ of up to $\unit[400]{\mu C/m^2}$ along the same axis\,\cite{Zvezdin2006}. This strong  magnetoelectric coupling sparked significant interest in \NdFeBO\ and immediately raised the question about its microscopic origin. Recent x-ray diffraction experiments already provided microscopic insight and implied that the finite \Pfe$_a$ is related to the magnetic field-induced CM order\,\cite{HamannBorrero2010,HamannBorrero2012HT}.
In principle, the strong magnetoelectric coupling discovered in previous studies should also enable to alter the magnetic order of  \NdFeBO\ by applying external electric fields. Yet no experimental studies of this effect have been available up to now. We therefore performed high-resolution magnetic Resonant X-ray Diffraction (RXD) in applied electric fields in order to close this gap.

The magnetic RXD has been performed at the resonant scattering beamline of the European Synchrotron Radiation Facility (ESRF) in Grenoble, France (formerly installed at ID20)\,\cite{Paolasini2007}. A liquid helium Orange cryostat equipped with an $E$-field stick --in-house built by one of us (C.M.)-- was mounted on a six circle diffractometer as described in Refs.\,\onlinecite{Fabrizi2009,Walker2011}, using an horizontal scattering geometry.
The sample with a thickness of $\unit[1.3]{mm}$ along the $a$-direction was mounted between two vertical electrodes, leaving an additional gap of $\unit[0.5]{mm}$ between the top electrode and the \NdFeBO\ crystal. By applying a voltage across the two electrodes, electric fields up to $E=$\Eval\  parallel to the crystallographic $a$ axis were generated.
%
We set the photon energy to the \NdLtwo\ edge at $\unit[6.726]{keV}$ in order to probe directly the magnetic order of the Nd-sublattice, which, in turn, also reflects the magnetic order of the Fe-sublattice\,\cite{HamannBorrero2012HT}.
For the $L$-scans shown in the following, the incoming light was horizontally ($\pi$) polarized and no polarization analyzer was used.
%


In Fig.\,\ref{fig:E-Leff}, $L$-scans through the magnetic superlattice reflections measured at $\unit[5.8]{K}$ are presented. Prior to these measurements the state of the sample was prepared by cooling down to $\unit[5.8]{K}$ without electric field. The $L$-scan obtained for this zero field cooled state is shown in  Fig.\,\ref{fig:E-Leff}\,(a). In agreement with previous results both the (0, 0, 4.5) of the CM phase as well as the ICM-satellites at (0,0,4.5$\pm \delta$) are observed, revealing the coexistence of the CM minority and the ICM majority phase under the present conditions. It is likely that the  observed residual intensity at  (0, 0, 4.5) is  due to pinned CM-domains that survive deep inside the ICM phase. This observation already indicates that both phases have very similar free energies and that small perturbations suffice to drive the magnetic system from one phase to the other.

The effect of an electric field on this phase coexistence can be observed in Fig.\,\ref{fig:E-Leff}\,(b).  Applying $E_a=$ \Eval\ results in a transfer of intensity from the ICM satellites to the central CM-reflection. The application of the electric field therefore stabilizes the CM-phase and increases its volume fraction at the expense of the ICM phase. This conclusion is verified by the fact, that the sum of the scattered intensities from the IC and the CM domains stays constant within 5\%. However, the transformation from the ICM to the CM phase is not complete within the studied $E$-field range, as incommensurate reflections retain a significant intensity. After the $E$-field is switched off, the zero field state is recovered, as demonstrated  in  Fig.\,\ref{fig:E-Leff}\,(c). Hence the effect of the applied $E$-field is reversible and our results do not show any significant hysteresis.

The $E$-field also has a corresponding effect on the full width at half maximum (FWHM) of the superlattice reflections, which measures the average correlation length 
of the CM and ICM ordered regions. In detail, the experimental resolution in reciprocal space, as determined using the neighboring (0,0,6) Bragg reflection, is $0.001\times2\pi/c$, which is equivalent to a real space resolution of about 760\,nm. At 5.8\,K and $E_a=0$\,V/m the width of the ICM-satellites is found to be resolution limited (FWHM=$0.001\times2\pi/c$), implying a correlation length $\xi_{ICM}>760$\,nm. Increasing the electric field to $E_a=$ \Eval\  causes a clear reduction of $\xi_{ICM}$ to about 476\,nm (FWHM=$0.0016\times2\pi/c$). Conversely, the correlation length of the CM-ordered regions increases with $E_a$ from $\xi_{CM}\simeq205$\,nm  at $E_a=0$\,V/m to $\xi_{CM}\simeq450$\,nm at $E_a=$ \Eval\ (FWHM decreases from $0.0037\times2\pi/c$ to $0.0017\times2\pi/c$). The comparable values for $\xi_{CM}$ and $\xi_{ICM}$ again show that phase transformation from ICM to CM is not complete at our highest electric field.


Note also that the observed changes as a function of $E$ cannot be caused by a heating of the sample. Heating would result in a strong change of $\delta$, which, however, is not observed. The changes in intensity and FWHM are therefore truly related to a change in the volume fractions occupied by the CM and ICM domains. The stabilization of the CM by an applied electric field further implies that the electric moment of this phase must be larger than that of the ICM phase. In fact, the comparison to macroscopic measurements indicates that only CM-phase is ferroelectric while the ICM phase does not posses a significant \Pfe\,\cite{Zvezdin2006JETPLett}. 

From the present experiments it is clear that an applied $E$-field has a large effect on the magnetic order of \NdFeBO.
The  effects observed here indeed go beyond earlier observations\,\cite{Fabrizi2009,Radaelli2008Y,Tokunaga2012,BodenthinPRL2008} in that the population of two distinct, but coexisting phases is altered by an electric field even far away from a magnetic transition. Interestingly, the transition between the ICM and the CM phases as a function of $E$ at constant temperature does not involve the continuous rotation of spins from the helical towards the collinear alignment ($\delta$ remains constant). This is in stark contrast to the behavior which is observed as a function of temperature in zero applied fields. 

When applying a magnetic field at a fixed temperature, we also find that $\delta$ remains constant, i.e., in this respect the effect of electric and magnetic fields is the same. Nonetheless, this observation is surprising and sets the present results apart from previous work: as we demonstrated earlier\,\cite{HamannBorrero2012HT}, upon field-cooling to a fixed  T$<$\Tic, the value of $\delta$ indeed does change with the magnetic field strength. This shows that the magnetic state at a given (B,T)-point in the phase diagram, depends on the path to this this point. 

\begin{figure}
\center{
   \includegraphics[width=0.9\columnwidth]{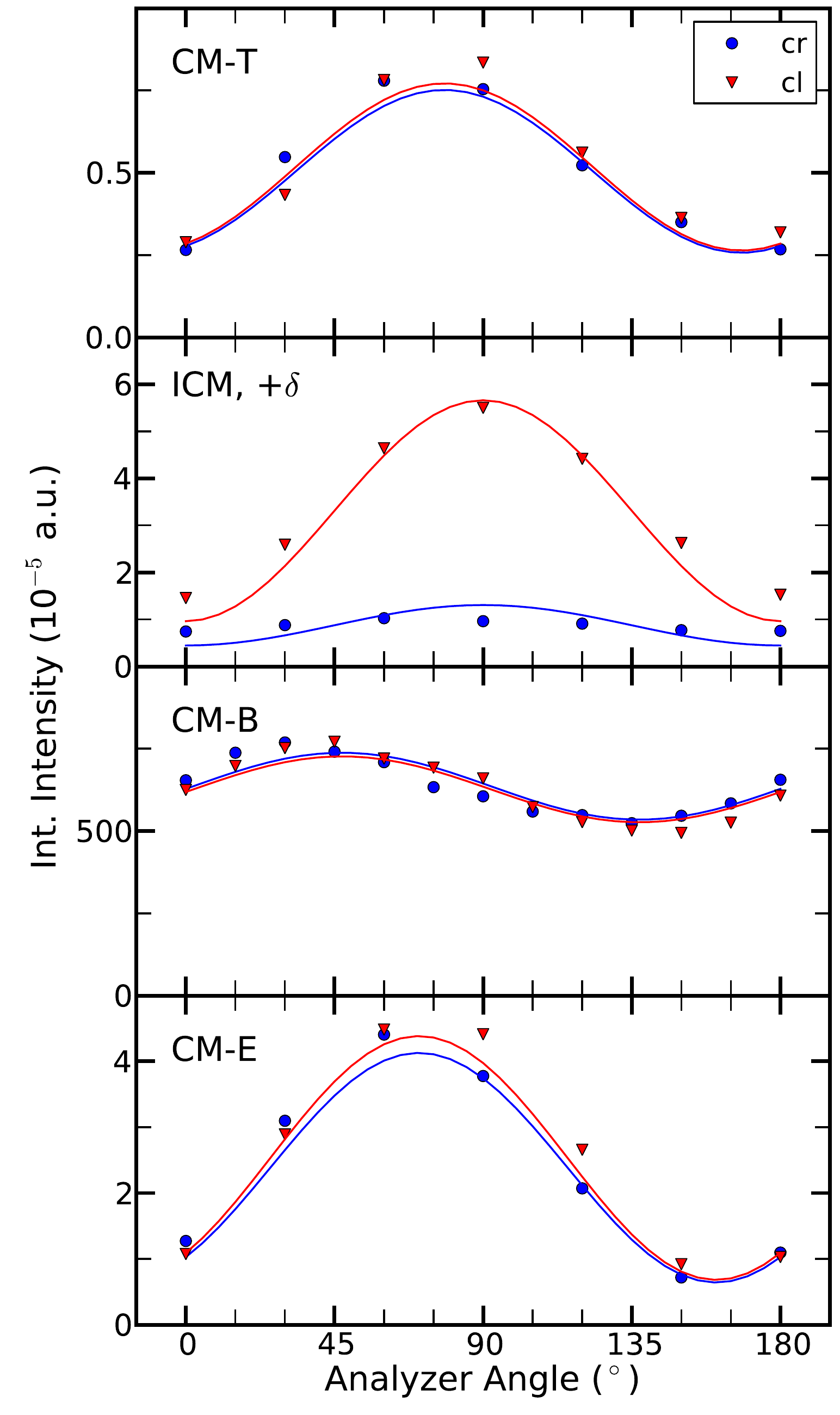}
 }
\caption{
(Color online) Comparison of experimental (symbols) and simulated  (lines) Stokes scans for circular right (\circr) and circular left (\cl)  polarized incoming photons.
The measured superlattice peaks are (0 0 7.5) and (0 0 7.5$+\delta$) for CM and ICM, respectively.
Temperature and fields for the shown measurements are:
CM-T: $\unit[20]{K}$, no field.
ICM: $\unit[5.8]{K}$, no field.
CM-B: $\unit[2]{K}$, $\unit[2]{T}$.
CM-E: $\unit[5.8]{K}$, \Eval.
}
\label{fig:Magnetic-configs}
\end{figure}


In order to get more microscopic information about the magnetic state under the various studied conditions, we performed a polarization analysis for circular right (\circr) and left (\cl) polarized incoming photons by means of so-called Stokes scans: 
for these scans, the polarization of the incoming and the detected photons are both controlled. The incoming beam polarization was set to \circr\ or \cl\ using a diamond phase plate, whereas the polarization of the scattered beam was monitored using a Cu220-analyzer. The polarization state of the incoming radiation, as characterized by the polarization parameters $(P_1, P_2, |P_3|)$\,\cite{Detlefs2012}, was  $(0.030, -0.006, 1.000)$ and $(0.042, -0.008, 0.999)$ for circular left and circular right, respectively, showing the high degree of circular polarization. 
During a Stokes scan, the polarization of the scattered beam in the plane perpendicular to its direction is measured as a function of the angle $\eta$, where $\eta=0^{\circ}$ corresponds to vertical 
and  $\eta=90^{\circ}$ to horizontal 
polarization.
We integrated rocking scans 
of the analyzer crystal at each $\eta$, to obtain the integrated intensity.
%
Examples and further details of the polarization analysis in magnetic RXD can be found in, e.\,g.,  Refs.\,\onlinecite{Detlefs2012,Mazzoli2007,Johnson2008}.

In this way, we investigated four different states: 
(i) the CM phase in zero fields at 20\,K (CM-T), (ii) zero field ICM at 5.8\,K (ICM), (iii)  the CM-phase induced by applied magnetic field (CM-B) at 2\,K and $B_a=2$\,T, as well as (iv) the CM-phase induced by applied electric field (CM-E).

The experimental data shown in Fig.\,\ref{fig:Magnetic-configs} was then compared to model calculations, in order to extract information about the microscopic magnetic configuration of the sample.
For modelling the magnetic RXD at the AFM superlattice positions we used the leading first order term of the spherical harmonic expansion of the scattering tensor, $F^{(1)}$\,\cite{Hannon1988,Haverkort2010}.
Since the magnetic moments are parallel to the $ab$-plane\,\cite{Fischer2006,Janoschek2010,HamannBorrero2012HT} and there is a 3-fold axis along $c$,  only one fundamental spectrum of $F^{(1)}$ enters, allowing us to apply the usual formalism given in Ref.\,\onlinecite{Hill1996}. Specifically, we use $F^{(1)}_{ij} \propto \epsilon_{ijk} m_k$, with $\epsilon_{ijk}$ and $m_k$ being the totally antisymmetric 3${}^{\mathrm{rd}}$ rank tensor and the magnetic moment direction, respectively. 
Employing these magnetic scattering tensors and the experimental geometry, the structure factor tensors for the different spin-configurations were calculated and compared to experiment.
%

For the CM-T phase we find that the polarization dependent magnetic RXD cannot be described by a mono-domain state. Instead, 
%
%
3 domains are necessary to reproduce the experimental data (top row in Fig.\,\ref{fig:Magnetic-configs}). While these measurements do not allow to determine the absolute spin directions of the different magnetic domains, we find that they must be rotated by 120$^{\circ}$ about c with respect to one another. This is expected because of the 3-fold axis of the unperturbed lattice structure (neglecting the small ferroelectric distortion).

The onset of the ICM order is signaled by the appearance of the IC satellites, which exhibit different resonant intensities for \circr\ and \cl\ (second panel from top in Fig.\,\ref{fig:Magnetic-configs}). This difference is due to the chirality of the magnetic helix. Specifically, our analysis yields the expected presence of  left and right handed domains with volume fractions of 0.70 and 0.30, respectively. 

The above findings for the CM-T and ICM phase are in good agreement with neutron scattering results\,\cite{Fischer2006,Janoschek2010}. It was also shown earlier that a magnetic field of 2\,T applied to the CM phase suffices to create a mono-domain CM-B state\,\cite{HamannBorrero2010,HamannBorrero2012HT}. Our analysis of the Stokes scans yields the same result, using a different and independent method. The above comparison to previous studies hence verifies our approach and shows that the full polarization control in magnetic RXD provides a powerful means to identify and characterize magnetic multi-domain states.



Turning to the CM-E, we find that 3 magnetic domains with a relative angle of 120$^{\circ}$  are necessary to model the experimental data,  very similar to the case of the CM-T state described above. We therefore find that our  maximum electric field of \Eval\/ 
does not create a mono-domain state with collinear magnetic order along one single direction. Instead still different types of domains are found to exist. This is in contrast to the mono-domain CM-B state described above.  


In summary, 
the key result of the present work is the observation of a delicate balance between two distinct magnetic phases, which reacts very sensitively to an applied electric field. More in detail, the present magnetic RXD experiments  reveal a complex microscopic state, where the coexisting CM- and ICM-regions also realize different domain-types, namely three 120$^{\circ}$ CM- and two left/right handed ICM-domains.
It is important to point out that frustrated magnetic couplings result in the competition of the CM and ICM order\,\cite{HamannBorrero2012HT}. Hence, this frustration also underlies the magnetic phase coexistence reported here, which occurs even deep inside the ICM phase and in the absence of a first order transition.
Triggering the magnetic phase coexistence in frustrated materials by electric fields therefore provides an alternative route towards new materials with a large magnetoelectric response.

\section{Acknowledgements}
We thank H.-C. Walker for her help with the experiments and technical support at ID20. We appreciate discussions with F. de Bergevin, C. Detlefs and L. Paolasini. 
S.P. and J.G. thank the DFG for the support through the Emmy Noether Program (Grant GE 1647/2-1). J.E.H.B. gratefully acknowledges the financial support of the DFG under grant HA6470/1-1. We also gratefully acknowledge the beamtime provision by the ESRF.

\end{document}